\documentstyle [prl,aps,multicol,amsfonts]{revtex}
\renewcommand{\narrowtext}{\begin{multicols}{2} \global\columnwidth20.5pc}
\renewcommand{\widetext}{\end{multicols} \global\columnwidth42.5pc}

\title{Termination of Multifractal behaviour for Critical Disordered Dirac
  Fermions}  

\author {J.-S. Caux$^1$, N. Taniguchi$^{1,2}$ and A. M. Tsvelik$^{1}$}

\address{$^1$ Department of Physics, University of Oxford, 1 Keble Road
  Oxford,OX1 3NP, UK\\ $^2$ Department of Physical Electronics, Hiroshima
  University, Higashi-Hiroshima 739, Japan}

\date{$Date:$ \today}

\begin{document} 

\bibliographystyle{prsty}

\draft \maketitle

\begin{abstract} 
  We consider Dirac fermions interacting with a disordered non-Abelian
  vector potential.  The exact solution is obtained through a special
  type of conformal field theory including logarithmic correlators,
  without resorting to the replica or supersymmetry approaches.  It is
  shown that the proper treatment of the conformal theory leads to a
  different multifractal scaling behaviour than initially expected.
  Moreover, the previous replica solution is found to be incorrect at
  the level of higher correlation functions.
\end{abstract}



\narrowtext

There exists a lot of evidence that the wavefunctions in disordered
systems exhibit multifractal behaviour near the
localization-delocalization transition (see for example
\cite{Wegner80,Aoki83,Janssen94,Huckestein95,Altshuler91}).  Multifractality manifests
itself in the anomalous scaling behaviour of moments of the local
density of states (LDOS) with system size.  This anomalous scaling is
described, for a d-dimensional system in a box of size $L$ with
short-distance cutoff $a$, by the spectrum $\tau^{*}(q)$ of exponents
defined as
\begin{equation}
  \left\langle \left[\rho(\bbox{r})\right]^{q}\right\rangle \sim
    (a/L)^{\tau^{*}(q)+d};
\label{size}
\end{equation}
Since $\tau^*(q)$ behaves linearly in $q$ for a simple fractal (for
example, a free wave in a box), the {\it nonlinear} dependence of
$\tau^{*}(q)$ signals the multifractality of the critical
wavefunctions.  Such a behaviour has been observed in numerical
calculations (for a review, see~\cite{Janssen94}) and a parabolic
dependence, $\tau^{*}(q) = (q - 1)(d-\alpha q)$, has been derived
explicitly in a few different perturbative approaches such as the
renormalization group (RG) one \cite{Wegner80,Altshuler91} and the
optimal fluctuation-like methods \cite{optimal}.
  
There are strong restrictions on the allowable behaviour of
$\tau^*(q)$ (see \cite{Janssen94}) which require among other things
the termination of the parabolic dependence for large enough $q$
(otherwise one ends up with unphysical arbitrarily negative exponents
in (\ref{size})).  One can imagine that this
termination should be achieved when all multiple-loop corrections are
taken into account in a perturbative scheme. To check whether this
hope is justified one inescapably needs to resort to nonperturbative
methods.

In this respect two-dimensional disordered systems take a special place
in investigating multifractality.  On the one hand, there are {\em
  critical} disordered systems that are known to exhibit
multifractality~\cite{Janssen94,Huckestein95}: familiar examples being
quantum Hall systems and disordered systems with spin-orbit
interactions.  On the other hand, in two dimensions one can study
critical behaviour using the powerful nonperturbative machinery of
conformal field theory (CFT) \cite{Belavin84}. It is then natural to
apply CFT to the problem of multifractality, and such attempts have
indeed been initiated in Ref.~\cite{Kogan96,Mudry96} by exploring
exactly solvable models of Dirac fermions (introduced in \cite{Ludwig94}
in connection with the Integer Quantum Hall Effect) in Abelian or
non-Abelian random vector fields.  However, the results obtained in
these papers are somewhat puzzling. In particular the solution to the
problem with $SU(N)$ random vector field contains an infinite set of
operators with {\it negative} conformal weights given by
\begin{equation}
h_q = \frac{q}{2} - \frac{N-1}{2N^2}(q^2 + Nq)
\label{hmudry}
\end{equation}
leading via RG-type arguments to the set of (negative for large
enough $q$) scaling exponents \cite{Mudry96}
\begin{equation}
\tau^*(q) = (q-1) \left(2 - \frac{N-1}{N^2}q \right) \label{taumudry} ,
\end{equation}
without explicit termination.  The absence of the
termination mechanism should be taken more seriously here than in the
case of RG calculations, since this CFT result is
nonperturbative and, if correct, exact.

We reveal in this letter the existence of a termination mechanism for
the non-Abelian case, emerging straightforwardly from the correct full
CFT treatment.  This mechanism invalidates the result (\ref{hmudry}).
For the sake of simplicity we illustrate this mechanism for the case $N
= 2$, in which $\tau^*(q)$ becomes negative for the smallest value of
$q$.  Generalisation to arbitrary $N$ is in principle possible.

The termination mechanism we present is different from the one in the
{\it Abelian} random vector potential case, which have been successfully
calculated in \cite{Chamon96,Castillo97} by mapping into a random energy
model or a type of Gaussian field theory.  Moreover, for {\it
  normalized} wavefuntions, the termination of the multifractal spectrum
can be expected to originate from the instability of the effective
Liouville field theory~\cite{Kogan96} for large $q$.

The non-Abelian model has the additional virtue of being exactly
solvable without using replicas or supersymmetry (SUSY)\cite{Efetov83}.
Therefore we also get a chance to independently check the results of
these two approaches \cite{Mudry96,Nersesyan94,Caux96} comparing them
with our exact results.  It is well known that the replica approach
fails to reproduce some (off-critical) nonperturbative results like the
level statistics in random matrix theory \cite{Verbaarschot85}.  For
critical disordered models one might expect more reliable results, since
criticality implies universality.  We here give an example demonstrating
that this is not the case. Our exact solution shows that the higher
correlation functions are identical with SUSY, but disagree with
replicas.

Another interesting aspect of our exact solution is in the nature of
the CFT involved.  One of the landmarks of conventional CFTs is the
power-law dependence of the physical correlators.  In fact, in many
instances, CFTs are identified by just such a property.  Recently,
however, it has been discovered that {\it logarithmic} dependence in
physical correlators can appear in certain models outside of the usual
class of so-called unitary minimal ones \cite{Gurarie93}.  Logarithms
are understood to be generated by degeneracies in the spectrum of
conformal dimensions of the theory: when two operators have dimensions
becoming degenerate, they metamorphose into a {\it logarithmic pair}
with unconventional correlators involving not only powers, but
logarithms.  The model we are considering, in fact, enters this class
of logarithmic CFTs.  Moreover, as we shall see, it is the presence of
such logarithmic operators that provides us with the solution to
the problems associated to negative scaling dimensions.

Although the details of the calculation are somewhat involved, the
conclusions can be relatively easily reached from our main result,
which is the fusion rule (\ref{opeMM}).  For the sake of clarity,
though, we briefly outline in what follows the calculation process
that leads to this result.  Interested readers can find more extensive
details in \cite{ourPRB}.

 We consider $N$ species of Dirac fermions living in a 2+1-dimensional
space and interacting through a disordered vector potential $A_{\mu}$
transforming like the adjoint of an $su(N)$ algebra ${\cal A}$, to which
they are coupled minimally.  
The disorder allows for hopping
between the different species.  Since the vector potential is
time-independent, different Matsubara frequencies do not couple, and can
be treated independently by a Euclidean two-dimensional theory with
explicit frequency dependence.  In fact, for a given realization of the
disorder, the partition function takes the form of the fermionic path integral
with the Dirac action
\begin{eqnarray}
S[\Psi, \omega, A_{\mu}] = \int d^2 x \bar{\Psi}(x) [{\bf I} ~\otimes \not
\! \partial -i \omega + i \not \!\! A] \Psi 
\end{eqnarray}
(since we are in a two-dimensional Euclidean space, we take the Pauli
matrices as Dirac $\gamma$ matrices).

Arbitrary products of disorder-dependent single-particle Green's functions
can then be calculated and averaged over the vector field distribution
functional
\begin{eqnarray}
P[A_{\mu}] = \frac{1}{\bar{g}} \int d^2 x \: \mbox{Tr}A_{\mu} (x) A_{\mu}
(x) \label{distribution} 
\end{eqnarray}
representing the usual $\delta$-correlated Gaussian white noise for the
random vector potential.

In the limits of infinite disorder strength $\bar{g} \to \infty$ and of
vanishing frequency $\omega \to 0$, the theory becomes conformally
invariant.  Correlators can then be calculated according to the
general principles behind CFT.

The derivation involves the following key moments.  We separate the
fermionic action into chiral parts, using holomorphic 
and antiholomorphic derivatives and fields $(2\partial = \partial_-,
2\bar{\partial} = \partial_+), ~ A_{\pm} = A_1 \pm i A_2$.  It is
important to note that now
$A_{\pm} \in su^C(N)$, the complex extension ${\cal A}^C$ of ${\cal
A}$.  We then parametrize the vector fields by fields $g_{\pm}$
belonging to the 
complex extension $G^C$ of the group $G = SU(N)$ as 
$A_{\pm} (x) = i \partial_{\pm} g_{\pm}(x) g_{\pm}^{-1}(x)$.
The reality condition $A_+^{\dagger} (x) = A_- (x)$ translates into
$g_+^{\dagger} (x) = g_- (x)$.  We will use the notation $g_+
(x) = g (x)$.  This reparametrization induces a
non trivial Jacobian in the path integral \cite{Mudry96,Bernard95}
\begin{eqnarray}
{\cal D} A_{\mu} = {\cal D} A_1 {\cal D} A_2 = {\cal D} G^C e^{2
N W [g^{\dagger} g]}  \label{jacobian}
\end{eqnarray}
where ${\cal D} G^C$ is the Haar measure over $SU^C(N)$, and
$W[g^{\dagger}g]$ is the 
Wess-Zumino-Novikov-Witten functional \cite{Knizhnik84}, here on level $k =
-2N$, for the field combination $g^{\dagger} g$.  This is a well-known
functional for which correlators are in principle known \cite{Knizhnik84}.

The next step is to decouple the fermions from the random potential 
by the transformations 
\begin{eqnarray}
\Psi_{\pm} (x) \to g_{\pm} (x) \Psi_{\pm}^{\prime} (x),  ~~~~~~
\Psi_{\pm}^{\dagger} (x) \to {\Psi_{\pm}^{\prime}}^{\dagger} (x)
g_{\pm}^{-1} (x)  \label{fermiondecoupling}
\end{eqnarray}
The Jacobian for this decoupling has the very important property of
being proportional to the partition function at fixed disorder
\cite{Bernard95}, thus
cancelling it when computing the correlations for a given realization.
This removes the need to invoke either the replica or SUSY
methods to perform explicitly the disorder averaging.

Let us then consider correlators of the local operator
\begin{eqnarray}
{\cal M}(z, \bar{z}) = \mbox{Tr} \bar{\Psi} \Psi
={\Psi_-^{\prime}}^{\dagger}_a h_{ab} {\Psi^{\prime}_+}_b +  
{\Psi_+^{\prime}}^{\dagger}_a h^{-1}_{ab} {\Psi^{\prime}_-}_b
\label{density} 
\end{eqnarray}
(in which $h_{ab} = [g^{\dagger}g]_{ab}$, with th $SU(N)$ indices)
which relates to the LDOS of the system and couples to the frequency
perturbation in the original action.  Notice that by doing this, we
explicitly sum over the $SU(N)$ indices: the product $g^{\dagger} g$
is invariant under left-multiplication of $g$ by $u \in SU(N)$.  This
is a crucial point: $g^{\dagger}g$ does not live on the complexified
group manifold of $SU^C(N)$, but rather on the coset space
$SU^C(N)/SU(N)$. The $SU(N)$ path-integration then simply factorizes
out of the effective generating functional.

In general, when one deals with CFTs, physical
states are associated to states in a highest-weight representation of
the Virasoro algebra \cite{Belavin84}.  Quantum mechanical commutation
rules are replaced in the radial quantization formalism by so-called
fusion rules, which state the short-distance singular behaviour of
products of operators in the complex plane defined by the Euclidean
spacetime variables $z = x_1 + ix_2, \bar{z} = x_1 -i x_2$.  Such
fusion rules are in principle completely obtainable from the
four-point correlation functions of the physical operators, for which
very powerful and extensive calculational methods are known.

One of the crucial aspects of the derivation is the proof that the
WZNW model of level $k= -2N$ on the coset space $SU^C(N)/SU(N)$
actually carries a representation of the Virasoro algebra for $SU(N)$
for the same (analytically continued to negative) level.  We refer the
reader to \cite{ourPRB} for an explicit proof of this statement.

The main procedures for obtaining the four-point coset correlators
\begin{eqnarray}
H = \langle h_{a_1 b_1} (z_1,\bar{z}_1) h_{b_2 a_2}^{-1} (z_2, \bar{z}_2)  
h_{a_3 b_3} (z_3, \bar{z}_3) h_{b_4 a_4}^{-1} (z_4, \bar{z}_4) \rangle
\end{eqnarray}
and thus the fusion rules, in the case of WZNW models is very clearly
set out in \cite{Knizhnik84}.  The ``conformal bootstrap'' provides
differential equations (the Knizhnik-Zamolodchikov equations) for the
so-called conformal blocks, whose solutions turn out to be
hypergeometric functions of the variable $z$.  
The peculiarity that we encounter in the ensuing CFT,
i.e. the one with $SU(2)$ Virasoro algebra on level $k= -4$, is that some
conformal blocks {\it have logarithmic behaviour}.  In fact, these
conformal blocks read
\begin{eqnarray}
&&\tilde{F}_1^a (z) = (1-z) F(3/2, 5/2; 2; z) \nonumber \\
&&\tilde{F}_1^b (z) = \frac{4}{\pi} \left[\frac{4/3}{z} + \tilde{F}_1^a
  (z) \ln z - 4/3 + (1-z) K_{11}(z) \right] \nonumber \\
&&\tilde{F}_2^a (z) = z F(3/2, 5/2; 3; z) \nonumber \\
&&\tilde{F}_2^b (z) = \frac{1}{2\pi} \left[\frac{16/3}{z} + \tilde{F}_2^a (z)
\ln z - 4/3 + z K_{12} (z) \right] \nonumber \\
\label{blocks}
\end{eqnarray}
where $F(a,b;c;z)$ is the hypergeometric function, and $K_{11},
K_{12}$ are some functions regular as $z \rightarrow 0$.

Enforcing the locality
conditions of quantum field theory then translates into ``gluing''
together the conformal blocks for the spacetime variables $z$ and
$\bar{z}$ in such a way that the combination is single-valued in the
complex plane defined by the variable $z$ (we refer the reader to
\cite{Dotsenko84} for the basic explanations).  The main point that we
want to stress here is that these gluing conditions in logarithmic
CFTs are not, as was once thought, unsolvable, but
rather a bit peculiar (for more details, see \cite{ourPRB}).
The crucial observation to make is that, as
we take $z \rightarrow e^{2\pi i} z$, only the combination $\ln
z\bar{z}$ remains invariant.  This dictates the proper gluing
procedure.

Leaving out the explicit details, we here simply provide
the final expression for the coset correlator ($z=
\frac{z_{12}z_{34}}{z_{13}z_{24}}, z_{ij} = z_i - z_j$):
\begin{eqnarray}
H = |z_{13}z_{24}|^{3/2} \sum_{i,j = 1,2} I_i \bar{I}_j H_{ij} (z,
\bar{z}) 
\label{cosetcor}
\end{eqnarray}
with
\begin{eqnarray}
H_{ij} (z, \bar{z}) = \alpha |z(1-z)|^{3/2} [\tilde{F}_i^a (z)
\tilde{F}_j^b (\bar{z}) + (a \leftrightarrow b)] \nonumber \\
I_1 = \delta_{a_1 a_2} \delta_{a_3 a_4} ~~~~~~ I_2 = \delta_{a_1 a_4}
\delta_{a_2 a_3} 
\end{eqnarray}
and for the resulting fusion rule between the coset operators (a
similar one has been obtained also in Ref.\cite{koglew}):
\begin{eqnarray}
\mbox{Tr}\; h (z_1, \bar{z}_1) h^{-1} (z_2, \bar{z}_2) \sim  
|z_{12}|^{3/2} \left[ \frac{4}{3}
[\frac{1}{z_{12}} A(z_2) + 
\right. \nonumber \\
\left. +\frac{1}{\bar{z}_{12}} \bar{A}(\bar{z}_2)] +
 \left[ 4 {\cal I} + 2 D 
(z_2, \bar{z}_2) + \ln |z_{12}| C(z_2, \bar{z}_2) \right] \right]
\label{opehh-1}
\end{eqnarray}
in which ${\cal I}$ is the identity operator.  The correlators of the
operators appearing in (\ref{opehh-1}) are
\begin{eqnarray}
&&\langle A(z_1) A(z_2) \rangle \sim z_{12}^2; \qquad \langle
\bar{A}(\bar{z}_1) \bar{A} (\bar{z}_2) \rangle \sim \bar{z}_{12}^2
\nonumber \\
&&\langle D (z_1, \bar{z}_1) D (z_2, \bar{z}_2) \rangle \sim - c_1 - \ln
|z_{12}| \nonumber \\
&&\langle D (z_1, \bar{z}_1) C (z_2, \bar{z}_2) \rangle \sim 1;
\nonumber\\ 
&&\langle C (z_1, \bar{z}_1) C (z_2, \bar{z}_2) \rangle = 0 \label{adcfusions}
\end{eqnarray}
where $c_i$ are some constants, unimportant for our purposes.

There are many remarks that we can make from this unconventional
Operator Product Expansion (OPE).
First and foremost, wee notice that the most relevant operators
appearing in the OPE (\ref{opehh-1}), $A(z)$ and $\bar{A}(\bar{z})$,
possess conformal weights $(-1, 0)$ and $(0, -1)$ respectively.
Usually, the fusion rules for WZNW models \cite{Knizhnik84} would
imply that the adjoint operator, whose conformal weights are $(-1,
-1)$, should appear in the OPE (\ref{opehh-1}).  But the term in the
four-point function pointing to the presence of such an operator, does
not appear since we are not allowed, by the requirement of single
valuedness, to multiply the logarithmic solutions in the holomorphic
and antiholomorphic sectors together.  This requirement cannot be seen
from the chiral conformal algebra studied in \cite{Mudry96}: it can
only come out of the solution for the full correlator that we have
obtained.

 The four-point ${\cal M}$ correlator can be calculated along the 
same lines as the coset correlator (\ref{cosetcor}) 
by  including the free fermion contributions and contracting the
indices appropriately.  
This four-point function allows us to extract the OPE of ${\cal M}$ with
itself, which in turn determines the scaling of the local moments.
  We find that the correct OPE reads \cite{ourPRB}:
\begin{equation}
{\cal M} (1) {\cal M}(2) \sim \frac{1}{|z_{12}|^{1/2}} \left[{\cal I} +
D(2) + \frac{1}{2} \ln |z_{12}| C(2) + ... \right]
\label{opeMM}
\end{equation}
where $\alpha$ is some constant.   The crucial fact is that the operator in
the symmetric representation does not appear here again, like in
(\ref{opehh-1}). $D$ and $C$ operators fuse as
(\ref{adcfusions}), whereas ${\cal M}, D$ and $C$ fuse as
\begin{eqnarray}
&&D(1) {\cal M}(2) \sim [\ln |z_{12}| + c_2] {\cal M}(2) + ... \nonumber \\
&&C(1) {\cal M}(2) \sim {\cal M}(2) + ... \label{fus}
\end{eqnarray}

 The fusion rules (\ref{opeMM},\ref{fus}) constitute the most
important result of our paper. The remarkable feature of these OPEs is
the appearance of logarithmic operators $D$ and $C$. Their origin lies
in logarithmic singularities present in multi-point correlation
functions.  These operators do not form the usual diagonalizable
representations of the Virasoro energy operator $L_0$ \cite{Gurarie93}
which is evident from their unconventional correlation functions
(\ref{adcfusions}).

The first comment that we can make about the OPE (\ref{opeMM}) is that it
invalidates the previous replica method treatment
\cite{Nersesyan94,Caux96}, for which the corresponding OPE read
\begin{eqnarray}
{\cal M}(1) {\cal M}(2) \sim \frac{1}{|z_{12}|^{1/2}} \left[ {\cal I}
+ z[ D(2) + \ln |z| C(2)] + \right. \nonumber \\
\left. +  \bar{z} [\bar{D}(2) + \ln |z| \bar{C}(2)]
+ ... \right]
\end{eqnarray}
in which $C, D, \bar{C}, \bar{D}$ were chiral logarithmic pairs with
conformal dimensions (1,0) and (0,1) respectively. This is different
from the pairs introduced in (\ref{adcfusions}) which have dimensions
(0,0).  Even though the two-point functions
coincide in both treatments, the higher-point functions are different.
Operator dimensions are calculated correctly by replicas, but
correlators of higher order are not.

Our result (\ref{opeMM}) will also lead to a markedly different
behaviour than the one associated to equation (\ref{taumudry}).  Since
the LDOS $\rho$ is related to the imaginary part of ${\cal M}$, its
local moments $\rho^q$ will scale at most like the {\it most relevant
part} of ${\cal M}^q$, which can be obtained by point-splitting from
(\ref{opeMM}).  We can draw the following comparison table between our
results for the conformal weights $h_q$ of the most relevant operator
contained in ${\cal M}^q$ as a
function of the power $q$, and the ones obtained in the previous
treatment \cite{Mudry96}:
\begin{equation}
\begin{array}{l|r|r|r|r|r}
~~~q & 1 & 2 & 3 & 4 & ...\\ \hline
h_q\; (\cite{Mudry96}) & ~1/8 & ~~-1 & -9/4 & ~~-4 & ...\\ 
h_q\; (\mbox{exact}) & 1/8 & 0 & 1/8 & 0 & ... \\ \hline
\end{array}
\end{equation}
The parabolic series of negative exponents is seen to be cut right
from the beginning for $SU(2)$. Note that logarithms are weight-zero
objects, so in fact they appear implicitly in our weights above for
all $q$.  The presence of these logarithmic prefactors makes the usual
treatment with $\tau^*$ exponents somewhat unsatisfactory, since it
can only describe pure power scaling and not the logarithmic
corrections to it that we have found.

In conclusion, we have shown that moments of the local density
operators in the problem of non-Abelian randomness obey different
scaling relations than expected \cite{Mudry96}, that do not fit in the
standard multifractal description.  The negative-dimensional operators
are suppressed by the presence of logarithmic operators in the
relevant OPEs.  Moreover, although replicas give correct primary field
scaling dimensions, they fail to reproduce the detailed form of the
correlators. Further details can be found in~\cite{ourPRB}.

 We are grateful to I. Kogan, I. Lerner, K. Efetov, A. Gogolin and
V. Kravtsov for interesting discussions and interest to the work. 
J.-S. C. acknowledges support from NSERC Canada, and from the Rhodes Trust.
N. T. acknowledges support from Japan Society for the Promotion of 
Science.

\widetext

\end{document}